# Large and tunable magnetoresistance in van der Waals Ferromagnet/Semiconductor junctions

Wenkai Zhu[1,2,9], Yingmei Zhu[2,3,9], Tong Zhou[4], Xianpeng Zhang[5], Hailong Lin[1,2], Qirui Cui[2,3], Faguang Yan[1], Ziao Wang[1,2], Yongcheng Deng[1], Hongxin Yang[2,3]✉, Lixia Zhao[1,6]✉, Igor Žutić[4]✉, Kirill D. Belashchenko[7]✉, Kaiyou Wang[1,2,8]✉

[1]State Key Laboratory of Superlattices and Microstructures, Institute of Semiconductors, Chinese Academy of Sciences, Beijing 100083, China

[2]Center of Materials Science and Optoelectronics Engineering, University of Chinese Academy of Sciences, Beijing 100049, China

[3]Key Laboratory of Magnetic Materials and Devices, Ningbo Institute of Materials Technology and Engineering, Chinese Academy of Sciences, Ningbo, Zhejiang 315201, China

[4]Department of Physics, University at Buffalo, State University of New York, Buffalo, New York 14260, USA

[5]Department of Physics, University of Basel, Basel, Basel-Stadt CH-4056, Switzerland

[6]Tiangong University, Tianjin 300387, China

[7]Department of Physics and Astronomy, Nebraska Center for Materials and Nanoscience, University of Nebraska-Lincoln, Lincoln, Nebraska 68588, USA

[8]Beijing Academy of Quantum Information Sciences, Beijing 100193, China

[9]These authors contributed equally: Wenkai Zhu, Yingmei Zhu.

✉e-mail: hongxin.yang@nimte.ac.cn; lxzhao@tiangong.edu.cn; zigor@buffalo.edu; belashchenko@unl.edu; kywang@semi.ac.cn

**Magnetic tunnel junctions (MTJs) with conventional bulk ferromagnets separated by a nonmagnetic insulating layer are key building blocks in spintronics for magnetic sensors and memory. A radically different approach of using atomically-thin van der Waals (vdW) materials in MTJs is expected to boost their figure of merit, the tunneling magnetoresistance (TMR), while relaxing the lattice-matching requirements from the epitaxial growth and supporting high-quality integration of dissimilar materials with atomically-sharp interfaces. We report TMR up to 192% at 10 K in all-vdW $Fe_3GeTe_2$/GaSe/$Fe_3GeTe_2$ MTJs. Remarkably, instead of the usual insulating spacer, this large TMR is realized with a vdW semiconductor GaSe. Integration of two-dimensional ferromagnets in semiconductor-based vdW junctions offers gate-tunability, bias dependence, magnetic proximity effects, and spin-dependent optical-selection rules. We demonstrate that not just the magnitude, but also the TMR sign is tuned by the applied bias or the semiconductor thickness, enabling modulation of highly spin-polarized carriers in vdW semiconductors.**

The traditional path to enhance the TMR[1,2] relies on carefully choosing insulators and common ferromagnets, such as MgO with Fe and Co[3,4]. As the MTJ size scales down, this approach poses many obstacles, from materials nonuniformity and deteriorating quality to the enhanced energy consumption and reduced stability[5]. The breakthroughs in vdW materials and the discovery of two-dimensional (2D) ferromagnets[6,7] suggest important opportunities to overcome these problems in all-vdW MTJs, where realizing a large TMR~200% could revolutionize magnetic random access memories (MRAM)[5].

MTJs with both conventional or vdW ferromagnets typically include an insulating spacer layer, which makes it easier to achieve high TMR compared to a semiconducting barrier. However, a realization of tunable spin-polarized transport in semiconductors is desirable for many emerging applications[1]. Conventional materials, such as δ-doped Fe/GaAs junctions, already provide a degree of tunability with bias-dependent sign reversal of interfacial spin polarization and TMR[8-12]. Because the observed spin-dependent signals in such systems are only modest, switching to 2D vdW materials could offer significant advantages by: (i) simultaneously increasing the TMR[13,14] and supporting highly spin-polarized carriers, and (ii) expanding the tunability of spin-dependent properties, as demonstrated, for example, through gate-tunable magnetic proximity effects in Co/graphene lateral spin valves[15].

In this work, we demonstrate a surprisingly large and tunable TMR of up to 192% in all-vdW $Fe_3GeTe_2$/GaSe/$Fe_3GeTe_2$ MTJs with a semiconducting GaSe spacer separating two $Fe_3GeTe_2$ (FGT) ferromagnets. This realization greatly expands materials design opportunities for semiconductor spintronics that are unavailable to MTJs with insulators[16,17], including applications in artificial neural networks[18,19] and spin-lasers[20]. To our knowledge, this TMR significantly exceeds the largest reported value in any MTJ with a semiconductor spacer[21].

The sketch of the MTJ devices is plotted in **Fig. 1a**, where the two FGT electrodes sandwich a GaSe layer and an hBN layer covers the whole junction to avoid oxidation. An out-of-plane magnetic field $B$ controls the magnetization alignment of the FGT electrodes. The devices (A, B, C, D, E, F and G) with different GaSe-layer thicknesses were fabricated using mechanical exfoliation and dry transfer method (See Methods), where the GaSe-layer thicknesses were determined by atomic force microscope (AFM) for device A, B, C, D, E, F and G are about 5.5, 6.5, 7.3, 8.2, 9.2, 10.0 and 15.6 nm, respectively (Supplementary Fig. 1). From the optical image of all the fabricated devices, the active junction overlap areas $A < 20$ $\mu m^2$, which are comparable to the typical magnetic domain sizes in FGT flakes[22].

We first investigate the current-voltage ($I$-$V_{bias}$) characteristics under applied perpendicular $B = -0.4$ T to ensure the parallel-magnetization configuration of the two FGT. To directly compare different devices, the normalized nonlinear current density-voltage $J$-$V_{bias}$ curves at 10 K for devices A to G are shown in **Fig. 1b**, where the nonlinear behavior of devices A and B with a thinner GaSe layer shows in a larger current range (inset of **Fig. 1b**). The nonlinear $J$-$V_{bias}$ characteristics reveal the existence of the tunneling-barrier between GaSe and FGTs. The band alignment at the FGT/GaSe interface obtained by fitting the Fowler-Nordheim tunneling plots[23] also reveals that the maximum tunneling-barrier of electrons is up to 0.83 eV (Supplementary Fig. 2).

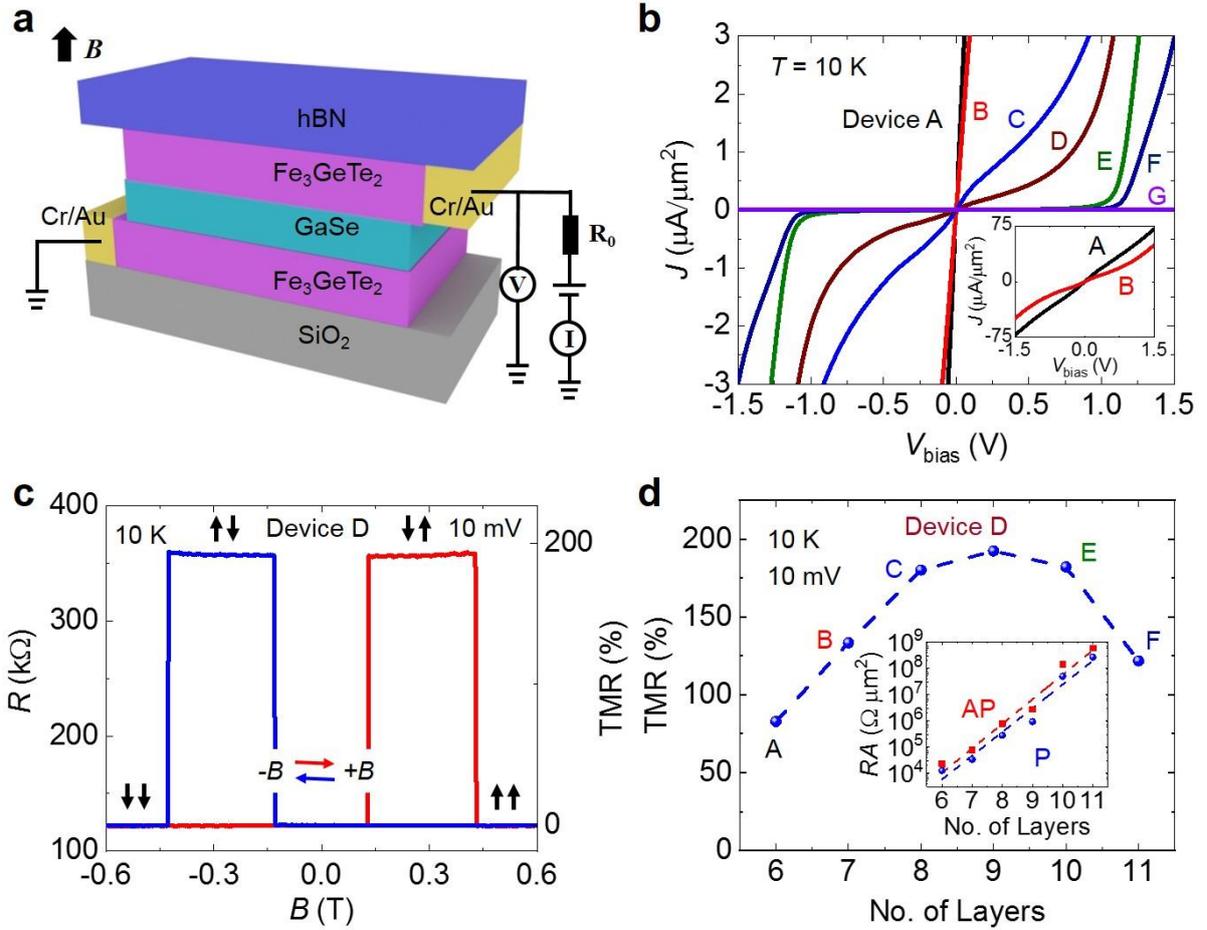

**Fig. 1 | Large TMR in the FGT/GaSe/FGT MTJ devices. a**, The schematic diagram of the device and magnetotransport setup. The magnetic field (***B***) is applied in out-of-plane direction. **b**, Current density $J$ versus applied bias $V_{bias}$ for the devices with the GaSe thickness ranging from 5.5 to 15.6 nm (devices A-G) in parallel-magnetic configuration. The inset shows the $J$-$V_{bias}$ curves of devices A and B in a larger bias range. **c**, Magnetic hysteresis of the resistance $R$ loop for device D at $V_{bias} = 10$ mV, and the corresponding TMR is ~192.4%. Red and blue horizontal arrows show the sweeping directions of ***B***. Black-vertical arrows denote the two FGTs' magnetization configurations. **d**, The measured maximum TMR ratios in the different devices at 10 mV. The inset shows the plotted zero-bias log($RA$) is nearly linear with the number of GaSe layers in both parallel and antiparallel states.

We next examine the TMR. Upon sweeping the out-of-plane ***B***, the devices A-F show two distinct parallel ($R_P$) and antiparallel resistance ($R_{AP}$) states (Supplementary Fig. 3). Among them, at bias of 10 mV, the $R_P$ and $R_{AP}$ of device D are 122.25 kΩ and 357.52 kΩ, respectively

(**Fig. 1c**). The corresponding TMR = $(R_{AP} - R_P)/R_P$ is 192.4%, which is the highest among the reported TMR devices with a semiconducting tunnel barrier[24,25]. The thickness of GaSe-layer dependence of the measured maximum TMR ratio is shown in **Fig. 1d**, which first increases from 83.1% (device A) and 133.5% (device B) and 180.2% (device C) to 192.4% (device D) and then decreases to 182.3% (device E) and 121.7% (device F), and finally vanishes (devices G) with increasing the thickness of GaSe-layer. The zero-bias resistance-area product ($RA$) of the devices for both the parallel and antiparallel states increases approximately exponentially with increasing the GaSe thickness, suggesting that the transport mechanism is dominated by tunneling (inset of **Fig. 1d**)[26]. The large TMR and nonlinear $J$-$V$ curve indicate that the GaSe spacer serves as a good tunneling-barrier. The maximum TMR at low bias is found in the device with GaSe of 8.2 nm, and the internal physical mechanism can be explained as follows. On one hand, the spin filtering effect of GaSe gets weaker when the thickness is reduced, resulting in the decrease of TMR[2]. On the other hand, with further increase of the spacer thickness, the TMR decreases and, eventually, vanishes in device G with 15.6-nm-thick GaSe, where the tunneling through extrinsic defects could become important and eventually exceed the spin-relaxation length of GaSe[27,28]. This experimental result suggests we can further improve TMR by optimizing the thickness of the semiconductor spacer layer.

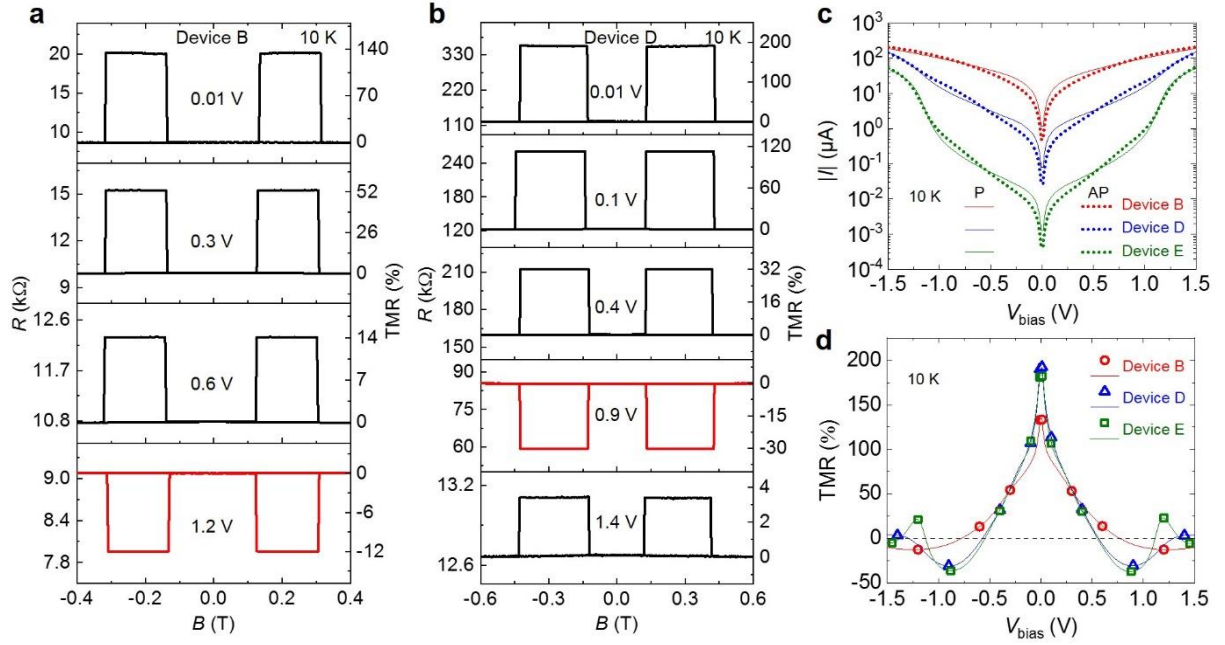

**Fig. 2 | The bias-dependent TMR of the devices. a-b**, The *R-B* curves at various positive bias for devices B and D. **c,** *I-V*$_{bias}$ curves of devices B, D and E in parallel and antiparallel states, respectively. **d**, The corresponding TMR as a function of *V*$_{bias}$. The hollow symbols are extracted from the *R-B* curves. The temperature is fixed at 10 K.

To investigate TMR(*V*$_{bias}$) for devices with different GaSe thickness, we measured the *R-B* curves. As shown in **Fig. 2a-b**, the positive TMRs decrease with *V*$_{bias}$. Negative TMRs of -12.3% and -30.5% for devices A and D are obtained at 1.2 V and 0.9 V, respectively. This salient sign inversion of TMR is found in all the devices A-F. The number of sign reversals of TMR can be tuned, from single to multiple, with increasing the GaSe thickness. To better understand the variation of TMR with bias, as shown in **Fig. 2c**, we measured the *I-V*$_{bias}$ curves of the devices in parallel and antiparallel states respectively. The nonlinear *I-V*$_{bias}$ curves for devices B, D and E show very different trends in parallel and antiparallel states, which allow us to derive bias-dependent TMR for these devices (**Fig. 2d**). The obtained TMR value matches well to that extracted from the *R-B* curves (**Fig. 2a-b** and Supplementary Figs. 4-5), indicating the influence of the Zeeman effect on TMR is negligible. The symmetric bias-dependent current and TMR suggest the symmetrical FGT/GaSe interfaces in these devices.

The devices A (Supplementary Fig. 6) and B show similar bias-dependent behavior of TMR. Specifically, as shown in **Fig. 2c**, for device B, the measured current for the parallel state is higher than that for the antiparallel state for $V_{bias} < 0.76$ V, leading to a positive TMR. However, beyond such $V_{bias}$, the measured current for the parallel state is lower than that for the antiparallel state, resulting in a negative TMR (**Fig. 2d**). With increasing the thickness of the GaSe spacer layer, we observed multiple sign changes of the TMR. As shown in **Fig. 2c**, the nonlinear $I$-$V_{bias}$ curves of device D in parallel and antiparallel states show two crossovers, and similar behavior is also observed in device C (Supplementary Fig. 7). Correspondingly, in **Fig. 2d**, the TMR of device D first decreases monotonously and changes sign of around 0.58 V, and again around 1.27 V as the bias increases. With further increasing the GaSe thickness, as shown in **Fig. 2c**, the nonlinear $I$-$V_{bias}$ curves reveal three crossovers for device E. In **Fig. 2d**, the TMR of device E first decreases monotonically and changes sign around 0.56 V, and then the TMR decreases in an oscillatory fashion as the bias increases. Similar three-sign changes of TMR behavior is also observed in device F (Supplementary Fig. 8).

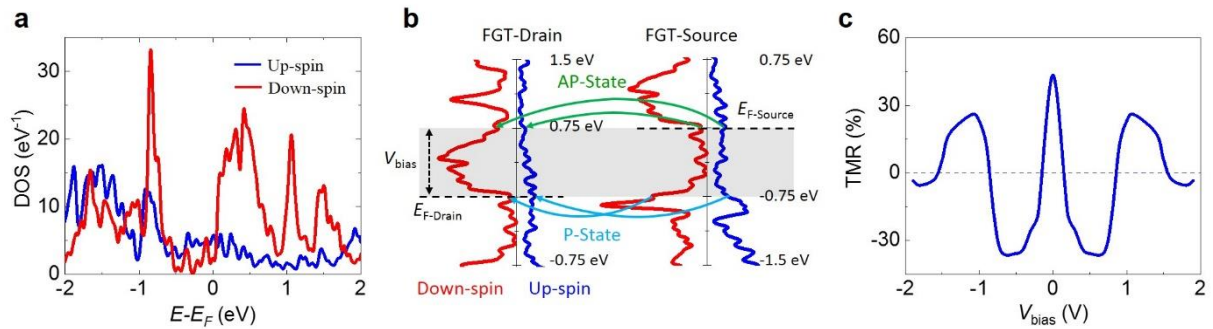

**Fig. 3 | The simulation of spin-resolved DOS of FGT and bias-dependent TMR calculated by elastic tunneling model. a**, The calculated spin-resolved DOS for both the up (blue line) and down (red) spins of the FGT in 3-layer-FGT/6-layer-GaSe/3-layer-FGT heterojunction. **b**, Schematic diagram of direct band-to-band spin-dependent tunneling under bias window $V_{bias}$ (shaded area). Light blue and green arrows represent the tunneling direction of the electrons in parallel and antiparallel states, respectively. **c**, The calculated TMR as a function of $V_{bias}$ by using the simple elastic tunneling formula.

To understand the bias-dependent magnetotransport, the spin-resolved density of states (DOS) of FGT in the 3-layer-FGT/6-layer-GaSe/3-layer-FGT heterojunction were obtained using the first-principles calculations (see Supplementary Fig. 9). The calculated spin-resolved

DOS of FGT electrode is shown in **Fig. 3a**. Assuming that tunneling is elastic and spin-conserving, with its tunneling probability independent of the initial and final states[29], the spin-dependent tunneling current at zero temperature can be expressed as[1,2,30] $I_\sigma \propto \int_{\mu_D}^{\mu_S} \rho_D^\sigma(E - \mu_D)\rho_S^\sigma(E - \mu_S)\, dE$, where the $\mu_{D(S)}$ and $\rho_{D(S)}^\sigma$ are the chemical potentials and spin-resolved DOS for the drain (or source) FGT, respectively ($\sigma$ is the spin index), and $\mu_D - \mu_{(S)} = eV_{\text{bias}}$ (**Fig. 3b**). The tunneling currents in parallel and antiparallel configurations can thus be expressed as $I_P \propto \int_{\mu_D}^{\mu_S} \left( \rho_D^\uparrow(E - \mu_D)\rho_S^\uparrow(E - \mu_S) + \rho_D^\downarrow(E - \mu_D)\rho_S^\downarrow(E - \mu_S) \right) dE$ and $I_{AP} \propto \int_{\mu_D}^{\mu_S} \left( \rho_D^\uparrow(E - \mu_D)\rho_S^\downarrow(E - \mu_S) + \rho_D^\downarrow(E - \mu_D)\rho_S^\uparrow(E - \mu_S) \right) dE$. The resulting bias-dependent TMR is shown in **Fig. 3c**. With increasing bias, the calculated TMR rapidly drops, changes sign, and then oscillates in qualitative agreement with the measurements for devices E and F. Multiple sign changes of TMR with increasing bias have also been predicted theoretically for Fe/MoS$_2$/Fe MTJ devices[31].

As the thickness of GaSe spacer decreases, the number of TMR reversals decreases to two in devices C and D and only one in devices A and B. The possible physical mechanism of the thickness-dependent TMR reversals is explained as follows.

The above calculation of the spin-dependent tunneling current assumed that the transmission probability is the same for all initial and final states. However, if tunneling is at least partially coherent, the transmission probability should be larger for states with transverse momenta close to those where the decay rate of the evanescent states in GaSe is the smallest. Because the band gap in GaSe is indirect, this may occur away from the Gamma point. The efficiency of this transverse-momentum filtering increases with increasing thickness of the tunnel barrier[30].

Because the interlayer dispersion of the vdW FGT states is weak, they are at a given transverse momentum essentially quantized. An energy isosurface of a lead thus consist of one or more 1D Fermi contours. Coherent tunneling is only possible at the intersections of the isosurfaces of the two leads corresponding to the same electrochemical potential. At large barrier thickness, a strong enhancement of the tunneling current should occur when such crossing points fall close to the points of the smallest decay rate in GaSe. If such matching condition is satisfied at the given bias for initial and final states of the same or opposite spin, positive or negative TMR is expected, respectively. The relative importance of such matching should increase at larger barrier thickness as long as coherent tunneling persists. Thus, coherent tunneling may explain why additional sign changes of TMR as a function of bias are observed at larger GaSe thicknesses.

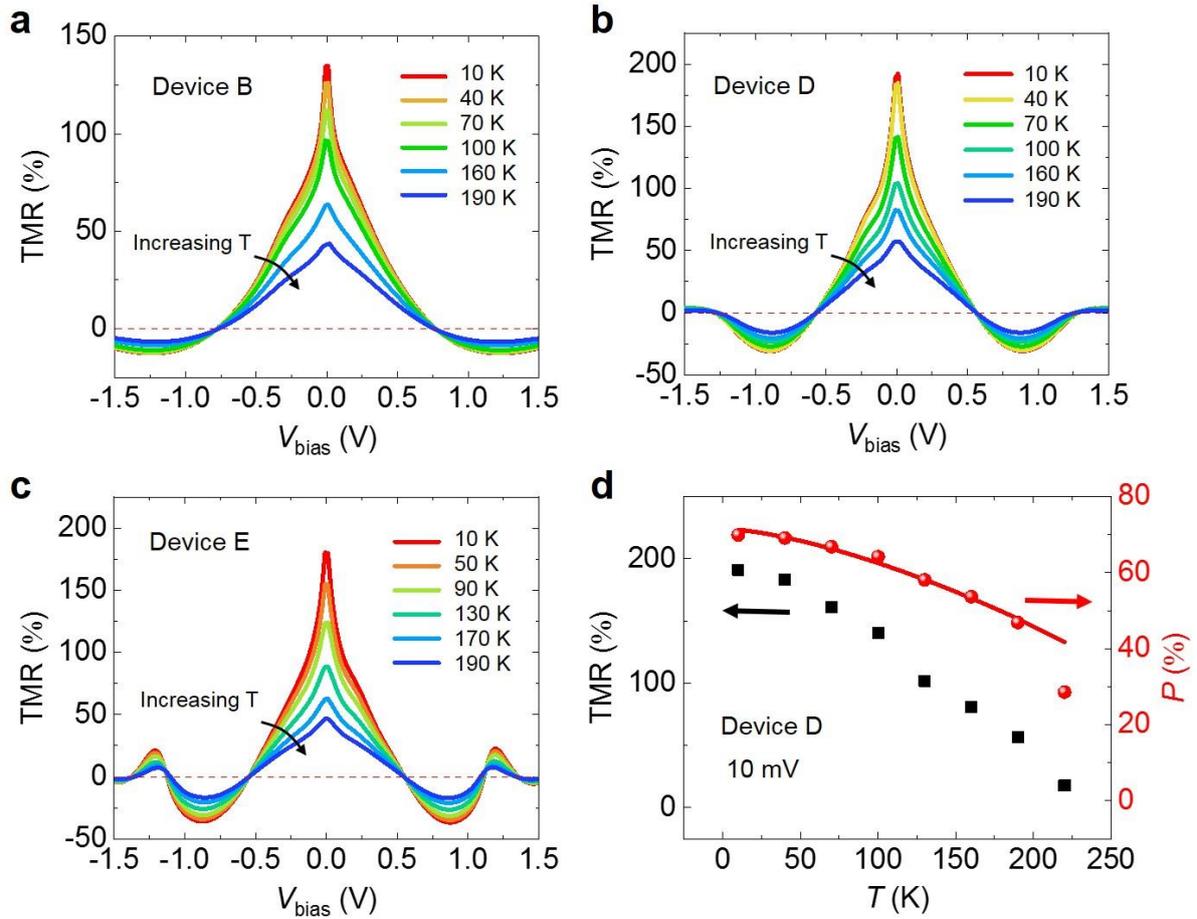

**Fig. 4 | The temperature-dependent TMR. a-c**, TMR ratios of devices B, D and E measured at temperatures from 10 K to 190 K, respectively. **e**, The TMR and spin polarization of device D as a function of temperature at bias of 10 mV. The red line shows the fitting data by the Bloch's low[1].

We further investigate the temperature-dependent TMR effect in our devices. The $I$-$V_{bias}$ curves of devices B, D and E in parallel and antiparallel states were measured at temperatures from 10 K to 190 K. The corresponding TMR is plotted in **Figs. 4a-c** for device B, D and E, respectively. The TMR-$V_{bias}$ curves in **Figs. 4a-c** all pass through the same zero at all temperatures, verifying that the bias-dependent TMR in the devices is dominated by tunneling[23]. The extracted TMR at different temperatures at 10 mV for device D (**Fig. 4d**) and device B and E (Supplementary Fig. 10) show a decreasing trend, which can be attributed to the decrease of spin polarization with temperature. When $V_{bias}$ is extremely small, approximately only the electrons at the $E_F$ take part in tunneling transport. For simplicity, assume the source and drain FGT electrodes have almost the same spin polarization. Then TMR can be defined as TMR = $2P^2/(1-P^2)$, where $P$ denote the spin polarization at the $E_F$ for the drain and source FGT electrode[1,2]. As shown in **Fig. 4d**, the $P$ decreases with temperature and the maximum $P$ at 10 K is up to 70%, which is about 4 times larger than that obtained in other 2D semiconductor-based MTJs[24]. The estimated temperature-dependence of the spin polarization can be fitted well by the Bloch's low, given by $P = P_0(1-\alpha T^{3/2})$, where $P_0$ is the spin polarization at 0 K, $\alpha$ is a materials-dependent constant[32]. The fitting value of $\alpha$ is 1.26-1.40×10$^{-4}$ K$^{-3/2}$, which is comparable to the previous reports[24,33].

Our presented results, which demonstrate large and tunable TMR, substantiate an ambitious vision where all-vdW MTJs could replace various charge-based memory applications[5], targeted to reach TMR~200% for commercial viability. Implementing such vdW MTJs is expected to rely on insulating 2D tunnel barrier[5], just as it was shown with h-BN barrier in an all-vdW MTJ with, at that time, the largest low-temperature TMR~160%[33] and also supported by a very recent report of TMR ~300%[34]. However, since we observe the desired large TMR values even with a semiconductor spacer, the prospect for all-vdW spintronics becomes considerably broader than just memory applications and the resulting large spin polarization and spin-orbit coupling opens opportunities beyond magnetoresistive effects. For example, the measured sign reversal of the TMR with applied bias is consistent with the reversal of the carrier spin polarization and could enable desirable polarization modulation[20]. Our findings could integrate semiconductor based optoelectronics, microelectronics and spintronics together, and could also relevant to emerging cryogenic applications where proximity-modified semiconductors and MTJs provide a platform for fault-tolerant quantum computing[35].

# References

1. Žutić, I., Fabian, J. & Das Sarma, S. Spintronics: Fundamentals and applications. *Rev. Mod. Phys.* **76**, 323-410 (2004).

2. Miao, G.-X., Münzenberg, M. & Moodera, J. S. Tunneling path toward spintronics. *Rep. Prog. Phys.* **74**, 036501 (2011).

3. Parkin, S. S. P. et al. Giant tunnelling magnetoresistance at room temperature with MgO (100) tunnel barriers. *Nat. Mater.* **3**, 862-867 (2004).

4. Yuasa, S. et al. Giant room-temperature magnetoresistance in single-crystal Fe/MgO/Fe magnetic tunnel junctions. *Nat. Mater.* **3**, 868-871 (2004).

5. Yang, H. et al. Two-dimensional materials prospects for non-volatile spintronic memories. *Nature* **606**, 663-673 (2022).

6. Gong, C. et al. Discovery of intrinsic ferromagnetism in two-dimensional van der Waals crystals. *Nature* **546**, 265-269 (2017).

7. Huang, B. et al. Layer-dependent ferromagnetism in a van der Waals crystal down to the monolayer limit. *Nature* **546**, 270-273 (2017).

8. Chantis, A. N. et al. Reversal of Spin Polarization in Fe/GaAs (001) Driven by Resonant Surface States: First-Principles Calculations. *Phys. Rev. Lett.* **99**, 196603 (2007).

9. Fujita, Y. et al. Nonmonotonic bias dependence of local spin accumulation signals in ferromagnet/semiconductor lateral spin-valve devices. *Phys. Rev. B* **100**, 024431 (2019).

10. Crooker, S. A. et al. Imaging Spin Transport in Lateral Ferromagnet/Semiconductor Structures. *Science* **309**, 2191-2195 (2005).

11. Lou, X. et al. Electrical detection of spin transport in lateral ferromagnet–semiconductor devices. *Nat. Phys.* **3**, 197-202 (2007).

12. Moser, J. et al. Bias dependent inversion of tunneling magnetoresistance in Fe/GaAs/Fe tunnel junctions. *Appl. Phys. Lett.* **89**, 162106 (2006).

13. Li, X. et al. Spin-Dependent Transport in van der Waals Magnetic Tunnel Junctions with $Fe_3GeTe_2$ Electrodes. *Nano Lett.* **19**, 5133-5139 (2019).

14. Zhang, L. et al. Perfect Spin Filtering Effect on $Fe_3GeTe_2$-Based Van der Waals

Magnetic Tunnel Junctions. *J. Phys. Chem. C* **124**, 27429-27435 (2020).

15. Xu, J. et al. Spin inversion in graphene spin valves by gate-tunable magnetic proximity effect at one-dimensional contacts. *Nat. Commun.* **9**, 2869 (2018).

16. Žutić, I. et al. Proximitized materials. *Mater. Today* **22**, 85-107 (2019).

17. Sierra, J. F. et al. Van der Waals heterostructures for spintronics and opto-spintronics. *Nat. Nanotechnol.* **16**, 856-868 (2021).

18. Sangwan, V. K. & Hersam, M. C. Neuromorphic nanoelectronic materials. *Nat. Nanotechnol.* **15**, 517-528 (2020).

19. Grollier, J. et al. Neuromorphic spintronics. *Nat. Electron.* **3**, 360-370 (2020).

20. Lindemann, M. et al. Ultrafast spin-lasers. *Nature* **568**, 212-215 (2019).

21. Matsuo, N. et al. High Magnetoresistance in Fully Epitaxial Magnetic Tunnel Junctions with a Semiconducting $GaO_x$ Tunnel Barrier. *Phys. Rev. Appl.* **6**, 034011 (2016).

22. Fei, Z. et al. Two-dimensional itinerant ferromagnetism in atomically thin $Fe_3GeTe_2$. *Nat. Mater.* **17**, 778-782 (2018).

23. Chiu, F.-C. A Review on Conduction Mechanisms in Dielectric Films. *Adv. Mater. Sci. Eng.* **2014**, 578168 (2014).

24. Lin, H. et al. Spin-Valve Effect in $Fe_3GeTe_2/MoS_2/Fe_3GeTe_2$ van der Waals Heterostructures. *ACS Appl. Mater. Inter.* **12**, 43921-43926 (2020).

25. Zhu, W. et al. Large Tunneling Magnetoresistance in van der Waals Ferromagnet/Semiconductor Heterojunctions. *Adv. Mater.* **33**, 2104658 (2021).

26. Tsunekawa, K. et al. Giant tunneling magnetoresistance effect in low-resistance CoFeB/MgO(001)/CoFeB magnetic tunnel junctions for read-head applications. *Appl. Phys. Lett.* **87**, 072503 (2005).

27. Jiang, X., Panchula, A. F. & Parkin, S. S. P. Magnetic tunnel junctions with ZnSe barriers. *Appl. Phys. Lett.* **83**, 5244-5246 (2003).

28. Takasuna, S. et al. Weak antilocalization induced by Rashba spin-orbit interaction in layered III-VI compound semiconductor GaSe thin films. *Phys. Rev. B* **96**, 161303 (2017).

29. Meservey, R. & Tedrow, P. M. Spin-polarized electron tunneling. *Phys. Rep.* **238**, 173-243 (1994).

30. Belashchenko, K. D. & Tsymbal, E. Y. Tunneling magnetoresistance: Theory, in *Spintronics Handbook: Spin Transport and Magnetism*, *2nd Edition*, edited by Tsymbal, E. Y. & Žutić, I. (CRC Press, Boca Raton, FL, 2019), pp. 525-558.

31. Dolui, K., Narayan, A., Rungger, I. & Sanvito, S. Efficient spin injection and giant magnetoresistance in Fe/MoS$_2$/Fe junctions. *Phys. Rev. B* **90**, 041401 (2014).

32. Evans, R. F. L., Atxitia, U. & Chantrell, R. W. Quantitative simulation of temperature-dependent magnetization dynamics and equilibrium properties of elemental ferromagnets. *Phys. Rev. B* **91**, 144425 (2015).

33. Wang, Z. et al. Tunneling Spin Valves Based on Fe$_3$GeTe$_2$/hBN/Fe$_3$GeTe$_2$ van der Waals Heterostructures. *Nano Lett.* **18**, 4303-4308 (2018).

34. Min, K.-H. et al. Tunable spin injection and detection across a van der Waals interface. *Nat. Mater.* (2022). https://doi.org/10.1038/s41563-022-01320-3.

35. Zhou, T. et al. Tunable magnetic textures in spin valves: From spintronics to Majorana bound states. *Phys. Rev. B* **99**, 134505 (2019).

## Methods

*Fabrication of the Fe$_3$GeTe$_2$/GaSe/Fe$_3$GeTe$_2$ MTJ devices.* The high-quality vdW bulk single-crystal FGT and hBN were purchased from HQ Graphene, while GaSe was purchased from 2D semiconductors, respectively. Firstly, a FGT flake was exfoliated onto polydimethylsiloxane (PDMS) stamps by adhesive tape. The stamps were adhered to a glass slide. Under optical microscope, the FGT flake with appropriate thickness and shape was chosen to transfer onto a 300 nm thick SiO$_2$/Si substrate by using a position-controllable dry transfer method[24]. Then, using the same method, a GaSe flake was transferred onto the FGT flake, followed by another thicker FGT flaker to fabricate a 2D heterojunction. To prevent the FGT from oxidation, a 20 nm-thick hBN layer was used to cap the whole heterostructure stack. Finally, the device was annealed at 120 ℃ for 10 minutes to remove the bubbles between the layers and ensure the close contact between the layers. Notably, the whole transfer processes were performed in nitrogen-filled glovebox with a concentration of less than 1 ppm of oxygen and water to ensure a clean interface. The source and drain electrode regions were pre-patterned by standard photolithography, and Cr/Au (10/40 nm) layers were deposited using an ultrahigh vacuum magnetron sputtering system, followed by a lift-off process. The thicknesses of both GaSe and FGT flakes were measured by using AFM (Bruker Multimode 8).

*Measurements of MR effect.* Electrical properties were measured using a semiconductor characterization system (Agilent Technology B1500A). All measurements were carried out in a Model CRX-VF Cryogenic Probe Station with a ±2.5 T out-of-plane vertical magnetic field.

*Ab Initio Calculations*: Our first-principles calculations are performed by the density functional theory (DFT) using the Vienna *ab initio* Simulation Package (VASP) code. The details are shown in the Supplementary Note 5.

## Data availability

The data that support the findings of this study are available from the corresponding author upon reasonable request.

## Acknowledgements

This work was financially supported by the National Key Research and Development Program of China (Grant Nos. 2017YFA0303400), the Beijing Natural Science Foundation Key Program (Grant No. Z190007), the National Natural Science Foundation of China (Grant Nos. 61774144, 11874059, 12174405), the Key Research Program of Frontier Sciences (Grant Nos. QYZDY-SSW-JSC020), the Strategic Priority Research Program of Chinese Academy of Sciences (Grant Nos. XDB44000000 and XDB28000000), the National Science Foundation, Grant No. ECCS-2130845, and the Air Force Office of Scientific Research, Grant No. FA9550-22-1-0349. K. D. B. is supported by the National Science Foundation through Grant No. DMR-1916275.

## Author contributions

K. W. conceived the work. W. Z. fabricated the devices, W. Z., F. Y., Y. D., H. L. and Z. W.

performed the experiments. W. Z., Y. Z., H. Y., L. Z., I. Ž., K. D. B. and K. W. analyzed the data. Y. Z., T. Z., X. Z., Q. C., H. Y., I. Ž., K. D. B. and K. W. carried out the modeling. W. Z., Y. Z. H. Y., L. Z., I. Ž., K. D. B., and K. W. wrote the manuscript. All authors discussed the results and commented on the manuscript.

## Competing interests

The authors declare no competing interests.

## Additional information

**Supplementary information** is available for this paper.

**Correspondence and requests for materials** should be addressed to Hongxin Yang, Lixia Zhao, Igor Žutić, Kirill D. Belashchenko or Kaiyou Wang.

# Supplementary Information

# Large and tunable magnetoresistance in van der Waals Ferromagnet/Semiconductor junctions

Wenkai Zhu[1,2,9], Yingmei Zhu[2,3,9], Tong Zhou[4], Xianpeng Zhang[5], Hailong Lin[1,2], Qirui Cui[2,3], Faguang Yan[1], Ziao Wang[1,2], Yongcheng Deng[1], Hongxin Yang[2,3]✉, Lixia Zhao[1,6]✉, Igor Žutić[4]✉, Kirill D. Belashchenko[7]✉, Kaiyou Wang[1,2,8]✉

[1]State Key Laboratory of Superlattices and Microstructures, Institute of Semiconductors, Chinese Academy of Sciences, Beijing 100083, China

[2]Center of Materials Science and Optoelectronics Engineering, University of Chinese Academy of Sciences, Beijing 100049, China

[3]Key Laboratory of Magnetic Materials and Devices, Ningbo Institute of Materials Technology and Engineering, Chinese Academy of Sciences, Ningbo, Zhejiang 315201, China

[4]Department of Physics, University at Buffalo, State University of New York, Buffalo, New York 14260, USA

[5]Department of Physics, University of Basel, Basel, Basel-Stadt CH-4056, Switzerland

[6]Tiangong University, Tianjin 300387, China

[7]Department of Physics and Astronomy, Nebraska Center for Materials and Nanoscience, University of Nebraska-Lincoln, Lincoln, Nebraska 68588, USA

[8]Beijing Academy of Quantum Information Sciences, Beijing 100193, China

[9]These authors contributed equally: Wenkai Zhu, Yingmei Zhu.

✉e-mail: hongxin.yang@nimte.ac.cn; lxzhao@tiangong.edu.cn; zigor@buffalo.edu; belashchenko@unl.edu; kywang@semi.ac.cn

**Table of Contents**

**Supplementary Note 1.** The thickness of the GaSe barrier layers

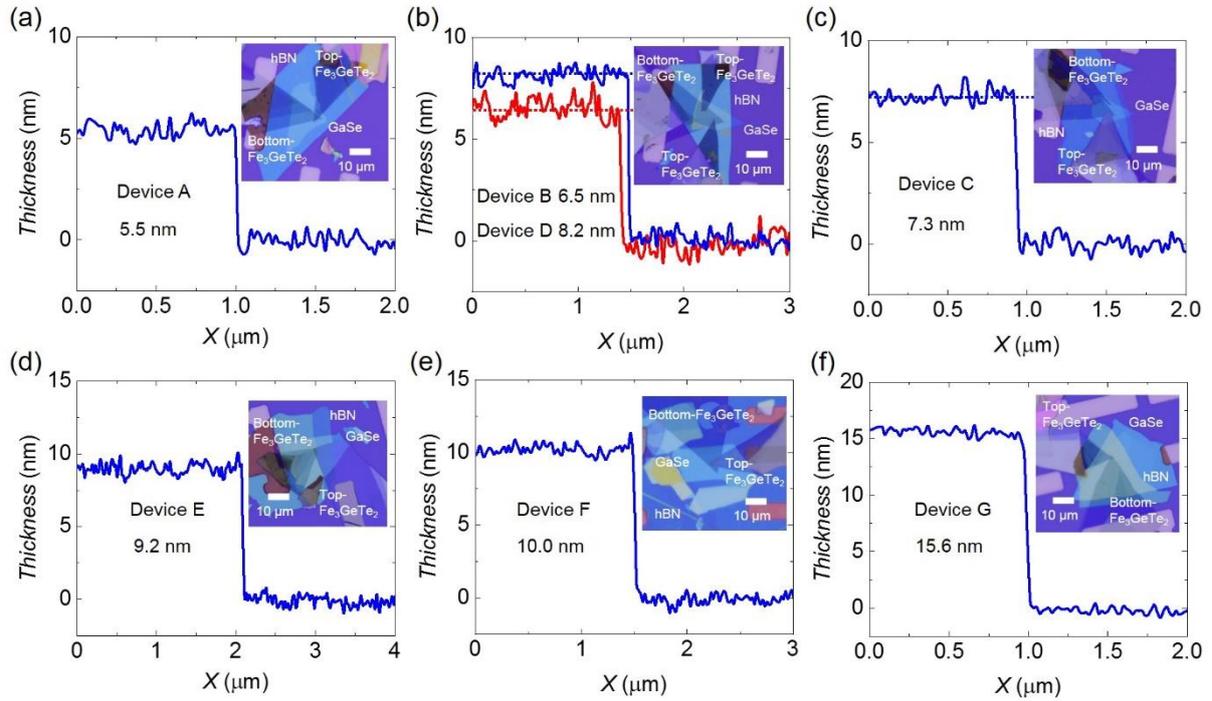

**Figure S1.** a-f) The scanning thickness of GaSe flakes for devices A-G by AFM measurements. The insets of the figures a-f) show the optical images of the devices A-G, respectively. The scale bar is 10 μm.

As shown in **Figure S1a-f**, the scanned thickness of the GaSe flakes by atomic force microscope (AFM) for devices A, B, C, D, E, F and G is about 5.5, 6.5, 7.3, 8.2, 9.2, 10.0 and 15.6 nm, respectively. Due to the thickness of monolayer of GaSe is about 0.93 nm[1], thus the devices A, B, C, D, E, F and G have 6-layers, 7-layers, 8-layers, 9-layers, 10-layers, 11-layers and 17-layers of GaSe monolayer, respectively. To have different coercive fields for the top and bottom electrodes, we select different thicknesses of FGT flakes[2,3].

**Supplementary Note 2.** The tunneling-barriers obtained from the Fowler-Nordheim tunneling plots

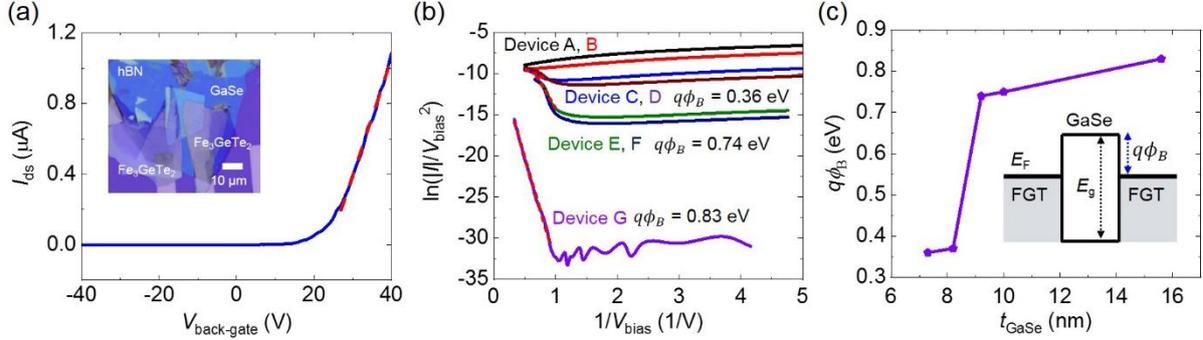

**Figure S2.** a) Transfer curve of a FGT/GaSe/FGT lateral FET device at 10 K. The inset shows the optical image of the FET device. b) F-N tunneling plots of $\ln(|I|/V_{bias}^2)$ vs $1/V_{bias}$ for the FGT/GaSe/FGT vertical MTJ devices at $T$ = 10 K. b) The extracted barrier heights for the devices with different GaSe thickness. The inset shows the energy band diagram of the device.

To determine the conductivity type of GaSe, we fabricated a lateral FGT/GaSe/FGT field effect transistor (FET) device and measured its transfer characteristic curve, as shown in **Figure S2a**. The field-effect mobility can be calculated using the formula: $\mu = (L/W)(d/\varepsilon_0\varepsilon_r V_{DS})(\Delta I_{DS}/\Delta V_{BG})$, where $L$ and $W$ are the length and width of the FET device, $\varepsilon_0$ (8.85 × 10$^{-12}$ Fm$^{-1}$) is the vacuum dielectric constant, and $d$ (300 nm) and $\varepsilon_r$ (3.9) are the thickness and dielectric constant of the SiO$_2$, respectively. Based on the transfer curve, the FET device presents an n-type conducting behavior with electron mobility of 2.9 cm$^2$V$^{-1}$s$^{-1}$. Unlike in common p-type semiconductor GaSe, the electronic conductivity in the FGT/GaSe/FGT FET device may be due to the charge transfer effect at the FGT/GaSe interfaces, which makes the Fermi level of GaSe move towards its conduction band[4].

We next analyze the barrier height of the FGT/GaSe/FGT MTJ devices by Fowler-Nordheim (F-N) tunneling formula. When a large bias is applied, the F-N tunneling becomes the dominant transport mechanism with the vertical charge flow at low temperature (10 K)[5,6]. The expression of the F-N tunneling current is $I = \frac{q^3 E^2 A}{8\pi h q \phi_B} \exp\left[\frac{-8\pi(2qm^*)^{1/2}}{3hE}\phi_B^{3/2}\right]$, where $q$ is the elementary charge, $E$ ($E = V_{bias}/t_{GaSe}$) is the external electric field, $A$ indicates the active overlap area, $m^*$ is the effective mass, $\phi_B$ is the effective tunneling-barrier height and $h$ is Planck's constant[5,6]. Hence, for F-N tunneling, a plot of $\ln(|I|/V_{bias}^2)$ versus $1/V_{bias}$ should be linear. In addition, the slope of the F-N tunneling plots can be expressed as a function of the effective mass and the effective tunneling-barrier height: $\text{slope} = -6.83 \times 10^9 \times t_{GaSe}\sqrt{\left(\frac{m^*}{m_0}\right)\phi_B^3}$. Extracting from the $J$-$V_{bias}$ data measured at 10 K (Figure 1b), the F-N tunneling plots of $\ln(|I|/V_{bias}^2)$ vs $1/V_{bias}$ are shown in the **Figure S2b**. As guided with red dotted lines, linear progressions of the F-N tunneling plots prove that F-N tunneling dominates charge transport at high $V_{bias}$ for the devices C, D, E, F and G[6,7]. GaSe shows an electron conducting behavior and $m^* \sim 0.1\ m_0$[8,9]. Therefore, the effective tunneling-barrier height for elections of devices C, D, E, F and G can be estimated as 0.36, 0.37, 0.74, 0.75 and 0.83 eV, respectively. The extracted barrier heights for devices with different GaSe thickness is shown in **Figure S2c**, where the tunneling-barrier height of the electron decrease with increasing the GaSe thickness. Since the band-gap of bulk GaSe is ~2 eV, the energy band diagram of the device is shown in the inset of **Figure S2c**. The band alignment with lager tunneling-barrier between the ferromagnetic metal and semiconductor is crucial for improving the TMR ratio of the devices; otherwise, the spin polarization would be lost dramatically due to the conductivity mismatch[10,11]. However, different from other devices, the F-N tunneling plots of the devices A and B increase with inverse bias, so the corresponding effective tunneling-barrier height cannot be calculated from the F-N tunneling formula.

**Supplementary Note 3.** The measured TMR in the MTJ devices with different GaSe-layer thickness

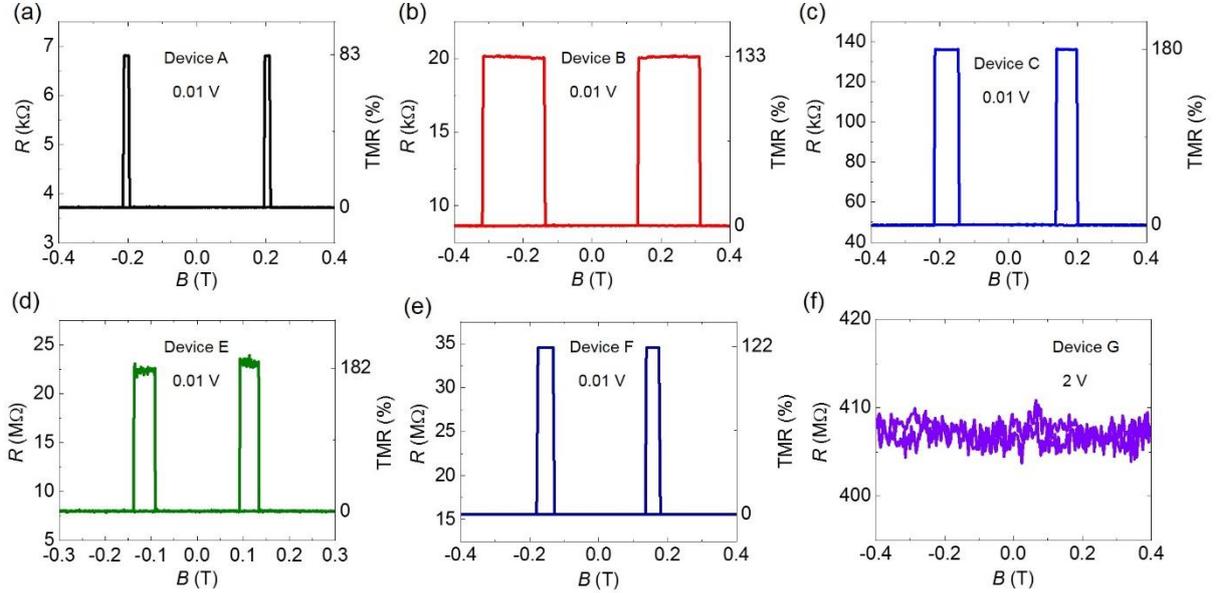

**Figure S3.** a) *R-B* curves of device A at 10 mV. The $R_P$ and $R_{AP}$ of device A are 3.72 kΩ and 6.81 kΩ, with TMR ~83.1%. b) *R-B* curves of device B at 10 mV. The $R_P$ and $R_{AP}$ of device B are 8.63 kΩ and 20.15 kΩ, with TMR ~133.5%. c) *R-B* curves of device C at 10 mV. The $R_P$ and $R_{AP}$ of device C are 48.58 kΩ and 136.11 kΩ, with TMR ~180.2%. d) *R-B* curves of device E at 10 mV. The $R_P$ and $R_{AP}$ of device E are 8.03 MΩ and 22.67 MΩ, with TMR ~182.3%. e) *R-B* curves of device F at 10 mV. The $R_P$ and $R_{AP}$ of device F are 15.59 MΩ and 34.57 MΩ, with TMR ~121.7%. f) *R-B* curves of device G under applied bias of 2 V. There is no TMR effect in device G. The temperature is fixed at 10 K.

**Supplementary Note 4.** The bias-dependent TMR in the MTJ devices

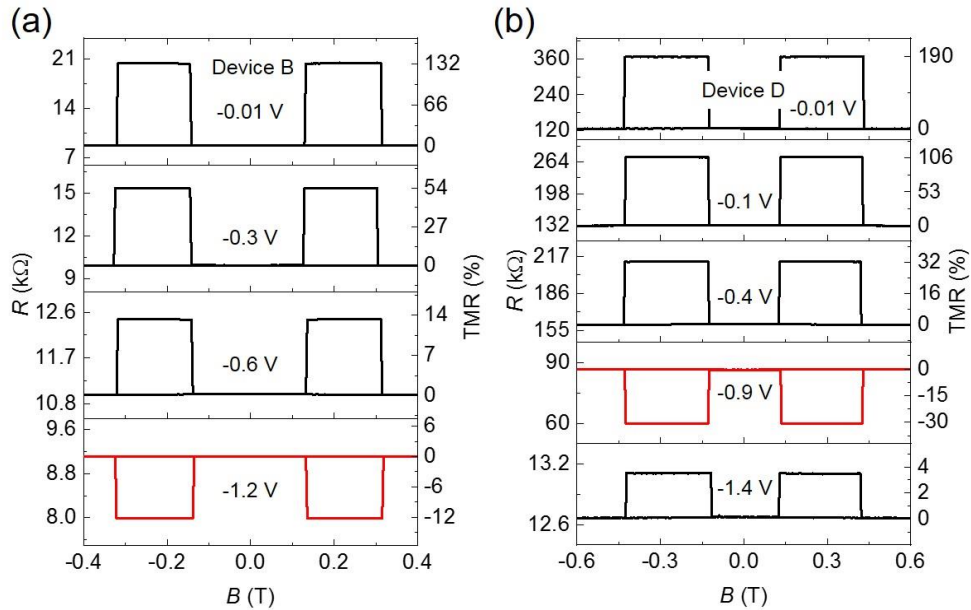

**Figure S4.** *R-B* curves and the corresponding TMR at various negative bias for devices B with 6.5-nm-thick GaSe (a) and D with 8.2-nm-thick GaSe (b) at 10 K.

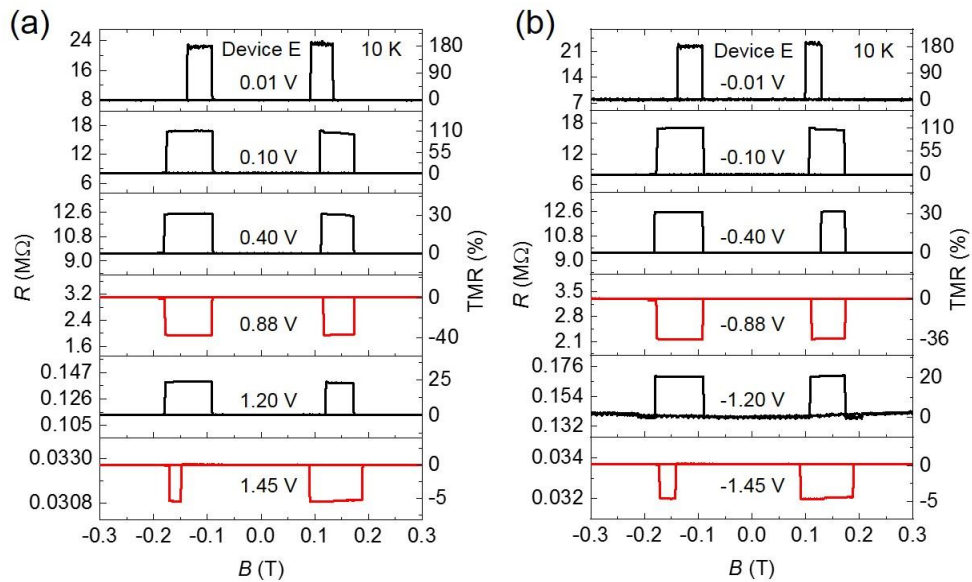

**Figure S5.** *R-B* curves and the corresponding TMR of device E with 9.2-nm-thick GaSe-layer at various positive bias a) and negative bias b). The temperature is fixed at 10 K.

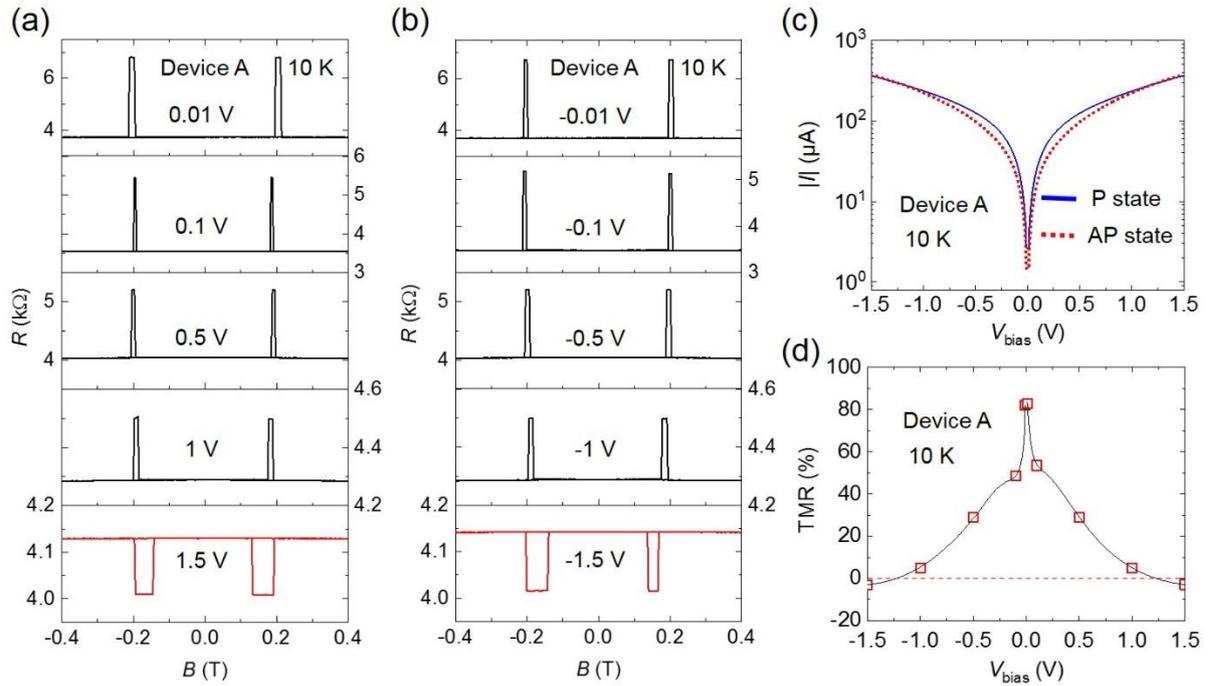

**Figure S6.** *R-B* curves of device A with 5.5-nm-thick GaSe-layer at various positive bias a) and negative bias b). c) *I-V*$_{bias}$ curves of device A measured in parallel and antiparallel magnetic configurations, respectively. d) TMR of device A as a function of *V*$_{bias}$, which decreases with the increase of bias and becomes negative when the bias exceeds 1.20 V. The temperature is fixed at 10 K.

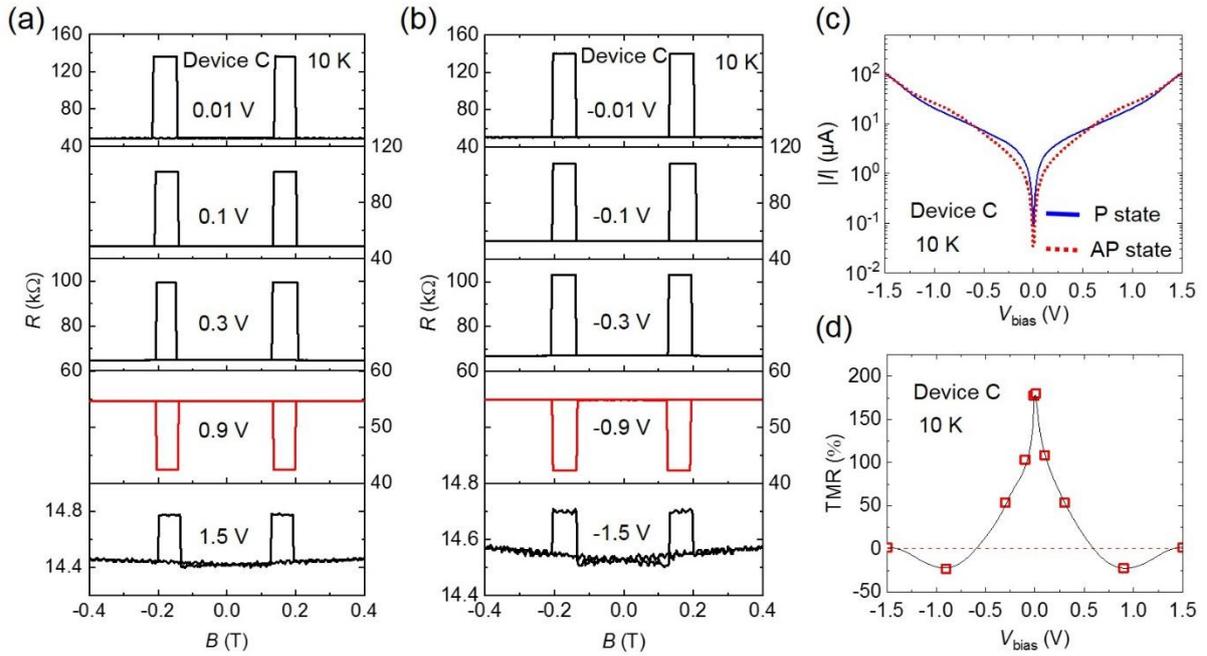

**Figure S7.** *R-B* curves of device C with 7.3-nm-thick GaSe-layer at various positive bias a) and negative bias b). c) *I-V*$_{bias}$ curves of device C measured in parallel and antiparallel magnetic configurations, respectively. d) TMR of device C as a function of *V*$_{bias}$, which decreases with the increase of bias and becomes negative when the bias exceeds 0.60 V and then back to positive value when the bias exceeds 1.43 V. The temperature is fixed at 10 K.

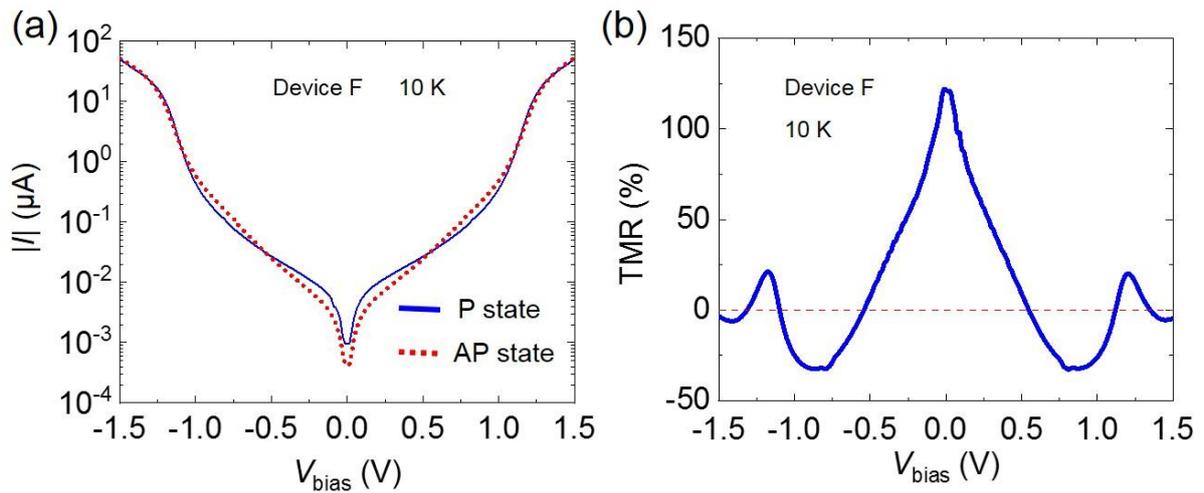

**Figure S8.** a) *I-V*$_{bias}$ curves of device F with 10-nm-thick GaSe-layer in parallel and antiparallel states, respectively. b) TMR versus *V*$_{bias}$ of device D. The temperature is *T* = 10 K.

**Supplementary Note 5.** The spin-resolved DOS simulated by DFT

*Ab Initio Calculations*: Our first-principles calculations are performed by the density functional theory (DFT) using the Vienna *ab initio* Simulation Package (VASP) code[12,13]. The exchange and correlation functionals are implemented by the generalized gradient approximation (GGA) of the Perdew-Burke-Ernzerhof functional[14]. A 2D FGT/GaSe/FGT van der Waals (vdW) heterostructure is constructed by sandwiching a 6-layer-GaSe between two 3-layer-FGTs. The lattice constant for pure FGT and GaSe is 3.99 and 3.81 Å, respectively. To obtain the correct spin polarization of FGT, the lattice constant of FGT/GaSe/FGT heterostructure is chosen as for the lattice of FGT, 3.99 Å, with a lattice mismatch of 4.7% compared to that of GaSe. Three different configurations between FGT and GaSe are considered: (i) Ga and Se atoms in GaSe-layer atop on Fe and Te atoms in FGT layer, respectively, (ii) Ga and Se atoms in GaSe-layer atop on Ge and Fe atoms in FGT layer, respectively, and (iii) Ga and Se atoms in GaSe-layer atop on Te and Ge atoms in FGT layer, respectively. The corresponding energy of these arrangements is -53.369, -53.440, and -53.448 eV, respectively. We find that the most stable configuration is (iii), as shown in Figure S9. The thickness of vacuum layer about 15 Å is set to avoid interactions between adjacent layers. Structures are fully relaxed until the force converged on each atom to less than $10^{-2}$ eV/Å. In all calculations, the plane-wave cutoff energy is set to 350 eV; Brillouin zone is sampled using Γ-centered 15×15×1 Monkhorst-pack *k* mesh[15]; and the vdW interaction are corrected by DFT-D3 Grimme method[16].

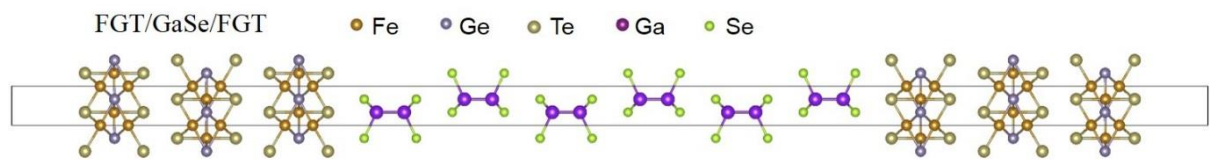

**Figure S9.** A side view of 3-layer-FGT/6-layer-GaSe/3-layer-FGT atomic structure used for the calculations.

**Supplementary Note 6.** The temperature-dependent TMR in the MTJ devices

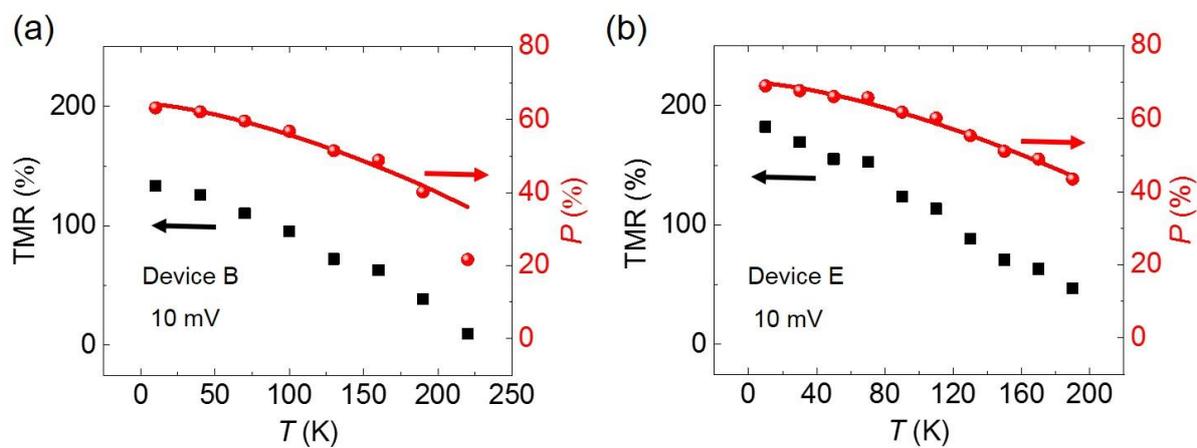

**Figure S10.** a) The extracted TMR and the corresponding $P$ of device B as a function of temperature at bias of 10 mV. The fitting value of $\alpha$ is $1.34\times10^{-4}$ K$^{-3/2}$. b) The extracted TMR and the corresponding $P$ of device E as a function of temperature at bias of 10 mV. The fitting value of $\alpha$ is $1.40\times10^{-4}$ K$^{-3/2}$.

References for Supplementary Information

1. Hu, P. et al. Synthesis of Few-Layer GaSe Nanosheets for High Performance Photodetectors. *ACS Nano* **6**, 5988-5994 (2012).

2. Skomski, R., Oepen, H. P. & Kirschner, J. Micromagnetics of ultrathin films with perpendicular magnetic anisotropy. *Phys. Rev. B* **58**, 3223-3227 (1998).

3. Tan, C. et al. Hard magnetic properties in nanoflake van der Waals $Fe_3GeTe_2$. *Nat. Commun.* **9**, 1554 (2018).

4. Chong, S. K. et al. Selective Growth of Two-Dimensional Heterostructures of Gallium Selenide on Monolayer Graphene and the Thickness Dependent p- and n-Type Nature. *ACS Appl. Nano Mater.* **1**, 3293-3302 (2018).

5. Chiu, F.-C. A Review on Conduction Mechanisms in Dielectric Films. *Adv. Mater. Sci. Eng.* **2014**, 578168 (2014).

6. Kim, S. et al. Highly Efficient Experimental Approach to Evaluate Metal to 2D Semiconductor Interfaces in Vertical Diodes with Asymmetric Metal Contacts. *ACS Appl. Mater. Inter.* **13**, 27705-27712 (2021).

7. Morgan, C. et al. Impact of Tunnel-Barrier Strength on Magnetoresistance in Carbon Nanotubes. *Phys. Rev. Appl.* **5**, 054010 (2016).

8. Manfredotti, C. et al. Electrical properties of p-type GaSe. *Il Nuovo Cimento B (1971-1996)* **39**, 257-268 (1977).

9. Fivaz, R. & Mooser, E. Mobility of Charge Carriers in Semiconducting Layer Structures. *Phys. Rev.* **163**, 743-755 (1967).

10. Rashba, E. I. Theory of electrical spin injection: Tunnel contacts as a solution of the conductivity mismatch problem. *Phys. Rev. B* **62**, R16267-R16270 (2000).

11. Dolui, K., Narayan, A., Rungger, I. & Sanvito, S. Efficient spin injection and giant magnetoresistance in $Fe/MoS_2/Fe$ junctions. *Phys. Rev. B* **90**, 041401 (2014).

12. Kresse, G. & Hafner, J. Ab initio molecular-dynamics simulation of the liquid-metal-amorphous-semiconductor transition in germanium. *Phys. Rev. B* **49**, 14251-14269 (1994).

13. Kresse, G. & Joubert, D. From ultrasoft pseudopotentials to the projector augmented-wave method. *Phys. Rev. B* **59**, 1758-1775 (1999).

14. Perdew, J. P., Burke, K. & Ernzerhof, M. Generalized Gradient Approximation Made Simple. *Phys. Rev. Lett.* **77**, 3865-3868 (1996).

15. Monkhorst, H. J. & Pack, J. D. Special points for Brillouin-zone integrations. *Phys. Rev. B* **13**, 5188-5192 (1976).

16. Grimme, S., Antony, J., Ehrlich, S. & Krieg, H. A consistent and accurate ab initio parametrization of density functional dispersion correction (DFT-D) for the 94 elements H-Pu. *J. Chem. Phys.* **132**, 154104 (2010).